\documentstyle[12pt]{article}
\def\ie{{\em i.e.}}
\def\eg{{\em e.g.}}

\def\beq{\begin{equation}}
\def\eeq{\end{equation}}

\catcode`\@=11 % This allows us to modify PLAIN macros.

\def\lsim{\mathrel{\mathpalette\@versim<}}
\def\gsim{\mathrel{\mathpalette\@versim>}}
\def\@versim#1#2{\vcenter{\offinterlineskip
    \ialign{$\m@th#1\hfil##\hfil$\crcr#2\crcr\sim\crcr } }}

\def\JL{J. L. Lopez}
\def\DVN{D. V. Nanopoulos}
\def\AZ{A. Zichichi}

\def\t1{{\tilde 1}}

\def\met{E\hskip-5.5pt/\hskip2pt}

\def\GeV{\,{\rm GeV}}
\def\TeV{\,{\rm TeV}}

\def\to{\rightarrow}
\def\pb{\,{\rm pb}}
\def\ipb{\,{\rm pb}^{-1}}
\def\ifb{\,{\rm fb}^{-1}}
\def \met	{/\!\!\!\!E_{T}}

\def\PLB#1#2#3{Phys. Lett. B {\bf#1} (19#2) #3}

\def\PRD#1#2#3{Phys. Rev. D {\bf#1} (19#2) #3}
\def\PRL#1#2#3{Phys. Rev. Lett. {\bf#1} (19#2) #3}

\def\IJMP#1#2#3{Int. J. Mod. Phys. A {\bf#1} (19#2) #3}
\def\TAMU#1{Texas A \& M University preprint CTP-TAMU-#1}

\begin{document}
% TH format
\begin{flushright}
\baselineskip=14pt
{CTP-TAMU-68/94}\\
{hep-ph/9412386}
\end{flushright}

\begin{center}
{\Large\bf  A survey of phenomenological constraints\\ on
supergravity models\\}
\vglue 0.5cm
{JORGE L. LOPEZ\\}
\vglue 0.25cm
{\em Department of Physics, Texas A\&M University\\}
{\em College Station, TX 77843--4242, USA\\}
E-mail: lopez@phys.tamu.edu
\end{center}

\vglue 0.25cm
\begin{abstract}
We advocate the study of supergravity models as well motivated few-parameter
low-energy supersymmetric models. In this context we survey a broad range
of phenomenological constraints and future tests, including present and
near-future Tevatron ($\tilde q,\tilde g,\chi^\pm_1,\tilde t_1$) and LEP
($h,\chi^\pm_1$) mass limits, collider indirect tests ($\Gamma_Z^{\rm
inv},R_b,m_t$), rare processes ($b\to s\gamma,(g-2)_\mu$), proton decay, and
dark matter (cosmology, direct and indirect detection).
\end{abstract}

\footnotetext{To appear in Proceedings of the Beyond the Standard Model IV
Conference, Lake Tahoe, California, December 13--18, 1994.}

\section{Why supergravity models?}
The most general ``minimal" low-energy supersymmetric model, \ie, the
Minimal Supersymmetric extension of the Standard Model (MSSM) has more than 20
free parameters, and therefore experimental tests and constraints are
impractical to implement. One needs further theoretical input to test/constrain
``sensible" models of low-energy supersymmetry. One promising avenue consists
of invoking models for physics at very high energies (for a recent review see
\cite{faessler}):
\begin{itemize}
\item Grand unification, provides gauge and Yukawa coupling relations, as well
as gaugino mass relations.
\item Supergravity, allows the calculation of the soft supersymmetry breaking
parameters in terms of the K\"ahler function ($K$) and the gauge kinetic
function ($f$). These constraints lead to four-parameter models.
\item Superstrings, provide specific forms for $K$ and $f$, and in principle
reduce the number of free parameters to {\em none}.
\end{itemize}

We can consider a ``generic" supergravity model described by the MSSM matter
content, gauge coupling unification at $\sim10^{16}\GeV$, a universal gaugino
mass ($m_{1/2}$), a universal scalar mass ($m_0$), a universal scalar coupling
($A$), and at low energies $\tan\beta$. The radiative electroweak breaking
mechanism allows the determination of $\mu,B$, and the parameter space is
four dimensional (plus the sign of $\mu$). Further constraints can reduce the
dimension of the parameter space, like Yukawa coupling unification which
determines $\tan\beta$ for a given value of $m_t$.

More ``modern" features have been recently incorporated in the study of
supergravity models, such as string unification at $\sim10^{18}\GeV$, simple
string models (\eg, $SU(5)\times U(1)$), and calculable values of the ratios
$\xi_0=m_0/m_{1/2}$ and $\xi_A=A/m_{1/2}$. In this case the four-parameter
models become two-parameter models. Moreover, in specific string models one can
also calculate the ratios of $\mu$ and $B$ to $m_{1/2}$, which lead to {\em
zero-parameter} models! It is also found that the soft supersymmetry
breaking terms are not necessarily universal.

At this point one can say that ``generic" supergravity models provide a
sensible ``standard" parametrization for comparisons of various
tests/constraints. However, ``modern" ingredients have subtle effects in the
sparticle spectrum, which sometimes produce new twists not present in
``generic" models. In fact, string model-building will likely soon provide a
new ``standard" parametrization, with novel/testable effects.

\section{Phenomenological constraints}
We do not consider theoretical constraints (\eg, gauge/Yukawa unification,
doublet-triplet splitting, fixed points, etc.). With few-parameter models in
hand, one can run the whole gamut of tests and produce an array of experimental
predictions. {\em Caution}: hidden assumptions make some constraints not as
strong as others (\eg, detection of dark matter assumes there is enough in the
halo to be detectable). Also, some constraints/tests are sensitive to a small
subset of sparticle spectrum, whereas others test the ``scale of
supersymmetry". As expected, the largest effects occur for the lightest
sparticle masses, but the ``reach" depends on the process and its experimental
sensitivity. We consider the following {\em not exhaustive} list of constraints
and tests:
\begin{itemize}
\item Collider mass bounds: LEP ($\chi^\pm_1,h,\tilde l,\tilde t_1$) and
Tevatron ($\tilde q,\tilde g,\chi^\pm_1,\tilde t_1$).
\item Collider indirect tests: LEP ($\Gamma^{\rm inv}_Z$, $R_b$, global fits)
and Tevatron ($t\to X$).
\item Rare processes: CLEO ($b\to s\gamma$) and Brookhaven ($(g-2)_\mu$)
\item Proton decay: SuperKamiokande ($p\to K^+\bar\nu$).
\item Dark matter: Cosmology (age and structure formation), direct dectection
in cryogenic detectors, and indirect detection in neutrino telescopes.
\end{itemize}

\subsection{Collider mass bounds}
\subsubsection{LEPI}
As is well known, all sparticles that couple to the $Z$ with unsuppressed
strength must have masses $\gsim{1\over2}M_Z$. This is the case for
$\chi^\pm_1,\tilde l^\pm,\tilde\nu,\tilde t_1$. The Higgs boson is produced
in the process $e^+e^-\to Z \to Z^* h,\ h\to 2j$, where
$\sigma_{\rm SUSY}=\sin^2(\alpha-\beta)\sigma_{\rm SM}$, and
$\sin^2(\alpha-\beta)_{\rm SUGRA}\approx1$. The latter is a consequence of
the supersymmetric decoupling phenomenon which is transmitted to the Higgs
sector via the radiative electroweak breaking mechanism \cite{LNPWZh}. Also,
for LEPI accessible mass scales, $B(H\to jj)_{\rm SM}\approx B(h\to jj)_{\rm
SUSY}$. These results imply $m_h\gsim62\GeV$. (In supergravity models $m_A>
m_h$, and thus the $hA$ mode is not accessible at LEPI.)
\subsubsection{LEPII}
$\bullet$ Gauginos: The mass relation $m_{\chi^\pm_1}\approx
m_{\chi^0_2}\approx 2m_{\chi^0_1}$ holds to various degrees of approximation in
this class of models. Charginos would be explored via
$e^+e^-\to\chi^+_1\chi^-_1\to 1l+2j$ (``mixed" mode) and, if the branching
ratios are not suppressed (this does happen though), the reach should be
$m_{\chi^\pm_1}\lsim{1\over2}\sqrt{s}$, as Fig.~\ref{ch-slep} shows \cite{One}.
The mode $e^+e^-\to\chi^0_1\chi^0_2,\chi^0_2\chi^0_2,\ \chi^0_2\to\chi^0_1+2j$
has a smaller cross section but a larger kinematical reach
$m_{\chi^\pm_1}\lsim{2\over3}\sqrt{s}$ \cite{LNPWZ}, and should be rather
background free.\\
$\bullet$ Sleptons: The processes of interest are
$e^+e^-\to\tilde e^+_{R,L}\tilde e^-_{R,L},
\tilde\mu^+_{R,L}\tilde\mu^-_{R,L},\tilde\tau^+_{1,2}\tilde\tau^-_{1,2}\to 2l$,
where $\tilde e_R,\tilde\mu_R,\tilde\tau_1$ are lightest and decay nearly
$\sim100\%$ to $l\chi^0_1$. The irreducible background $\sigma(WW\to
2l)=0.9\pb$ limits the reach somewhat, as Fig.~\ref{ch-slep} shows
\cite{One}.\\
$\bullet$ Higgs: $e^+e^-\to Zh$, $h\to b\bar b$, the novelty is that
$h\to\chi^0_1\chi^0_1$ may spoil the $b\bar b$ signal (for
$m_h>2m_{\chi^0_1}\approx m_{\chi^\pm_1}$) \cite{LNPWZ}. The reach is estimated
to be $m_h\lsim\sqrt{s}-95$ \cite{Sopczak}.

\subsubsection{Tevatron}
$\bullet$ Missing $E_T$: $p\bar p\to\tilde g\tilde g,\tilde g\tilde q,\tilde
q\tilde q\to{\met}+l's+j's$ is near the ``kinematical" limit. The latest bounds
are $m_{\tilde q}\sim m_{\tilde g}\sim200\GeV$ \cite{D0}. {\em Caution}: limits
apply only to the specific choice of supersymmetry parameters used in the
analysis.\\
$\bullet$ dileptons/trileptons: estimated sensitivities for
Run-IB/Main-Injector \cite{D0note,DiTevatron}.\\
\begin{tabular}{|l|c|c|} \hline
process&$100\ipb$&$1\ifb$\\ \hline
$p\bar p\to \chi^0_2\chi^\pm_1\to 3l$&$0.4\pb$&$0.07\pb$\\
$p\bar p\to\chi^+_1\chi^-_1\to 2l$&$1\pb$&$0.3\pb$\\ \hline
\end{tabular}
For light sleptons, $B(\chi^\pm_1\to l^\pm)$ may be enhanced but at the same
time $B(\chi^0_2\to 2l$) may be suppressed \cite{LNWZ}. When either
$\chi^0_2\to\chi^0_1 Z,\chi^0_1 h$ are allowed, the trilepton signal becomes
unobservable. As Fig.~\ref{di-tri} shows \cite{di-tri}, the two signals can
have an important complementary
role when the neutralino branching fraction is suppressed. In
$100\ipb\,(1\ifb)$ the reach in chargino masses using both signals can be as
high as $100\,(150)\GeV$.  \\
$\bullet$ Light top-squark: The process
$p\bar p\to\tilde t_1\tilde t_1\to(\chi^+_1 b)(\chi^-_1\bar b)\to 2l+2j$ has
an estimated sensitivity of $m_{\tilde t_1}\sim(100-130)\GeV$ (in $100\ipb$)
\cite{BST}.

\begin{figure}[p]
\vspace{3in}
\includegraphics{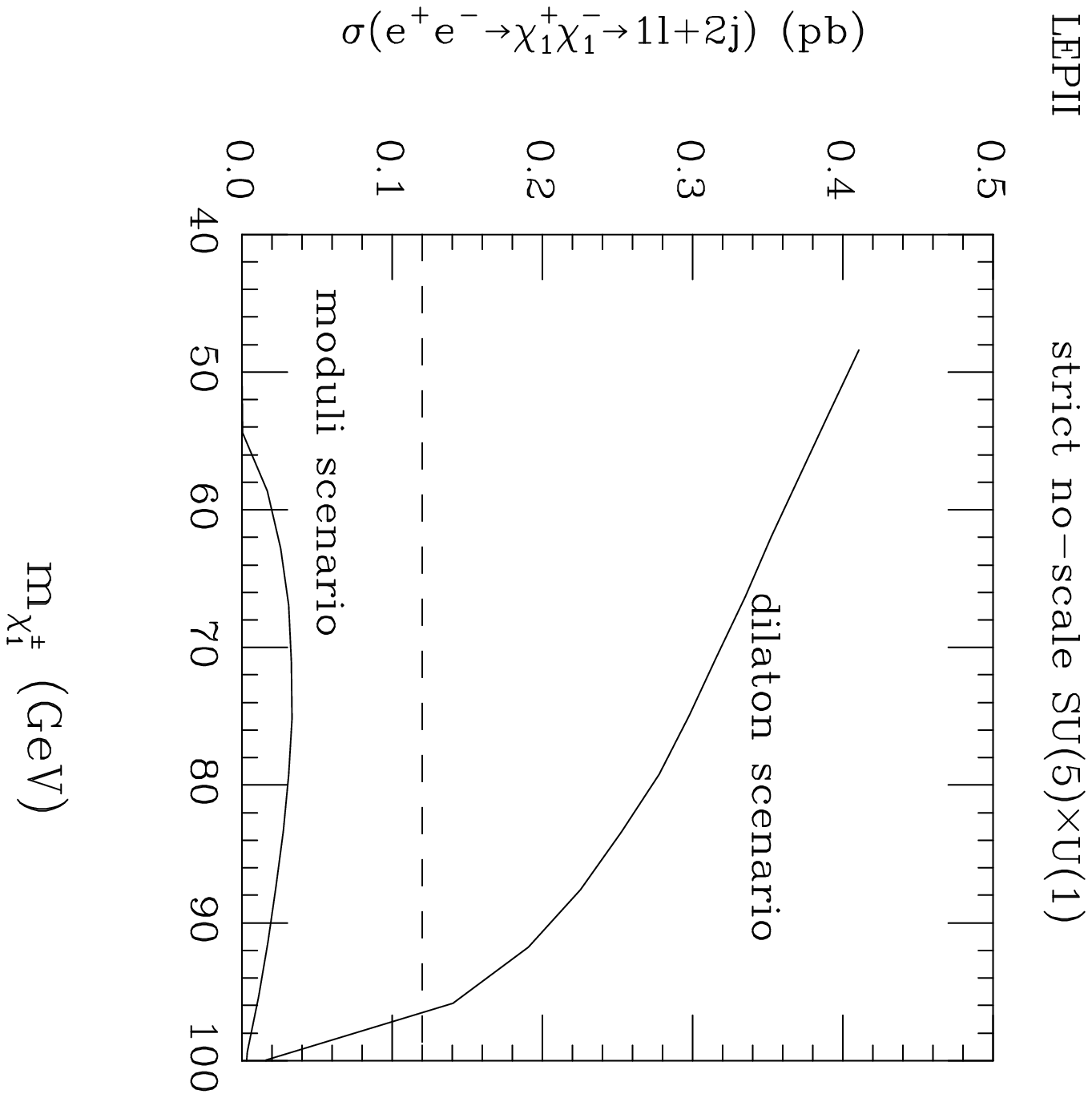}
\includegraphics{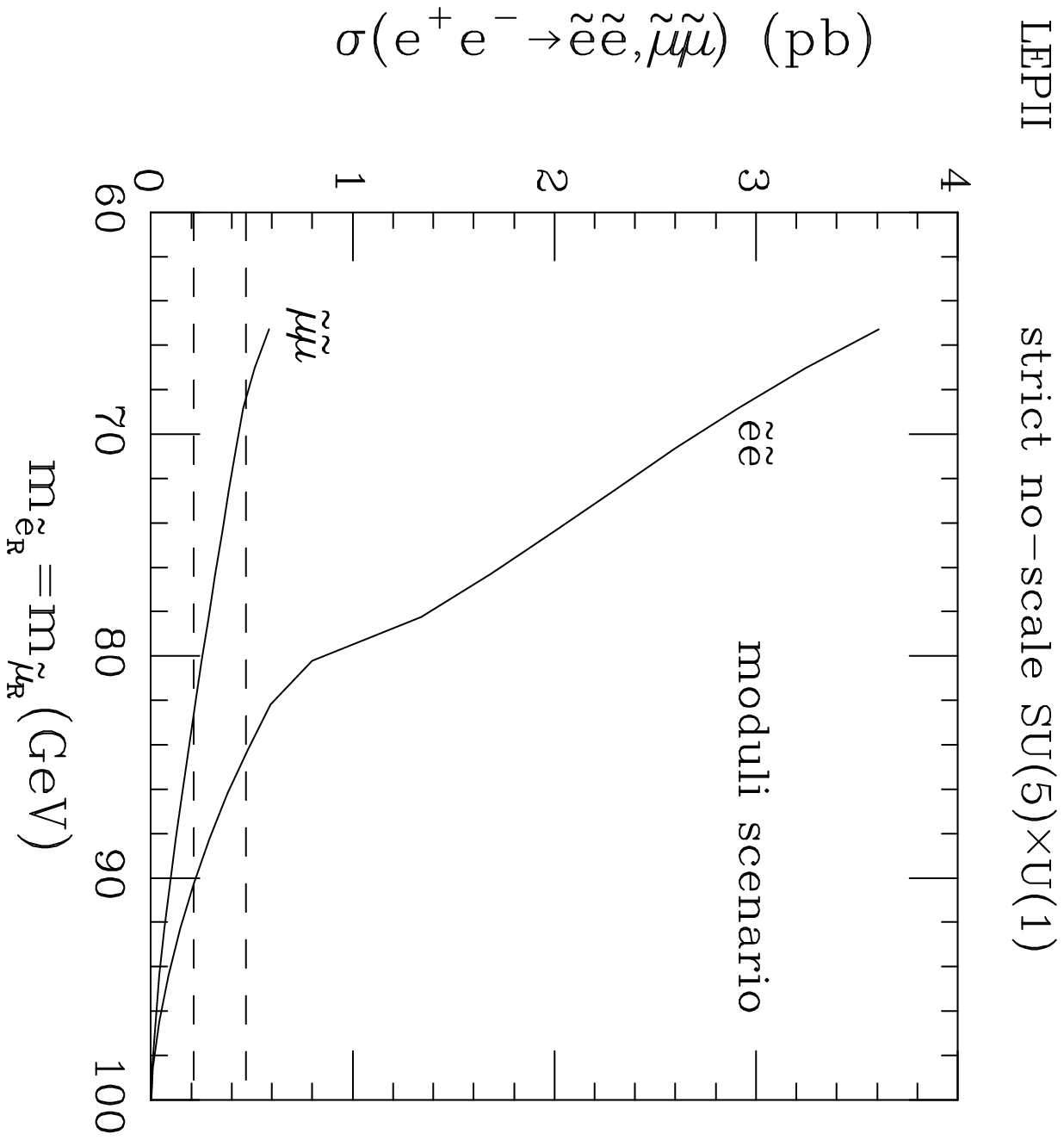}
\vspace{-0.5cm}
\caption{Chargino and slepton production at LEPII in one-parameter $SU(5)\times
U(1)$ supergravity. The dashed lines indicate the expected experimental
sensitivity.}
\label{ch-slep}
\vspace{5.0in}
\includegraphics{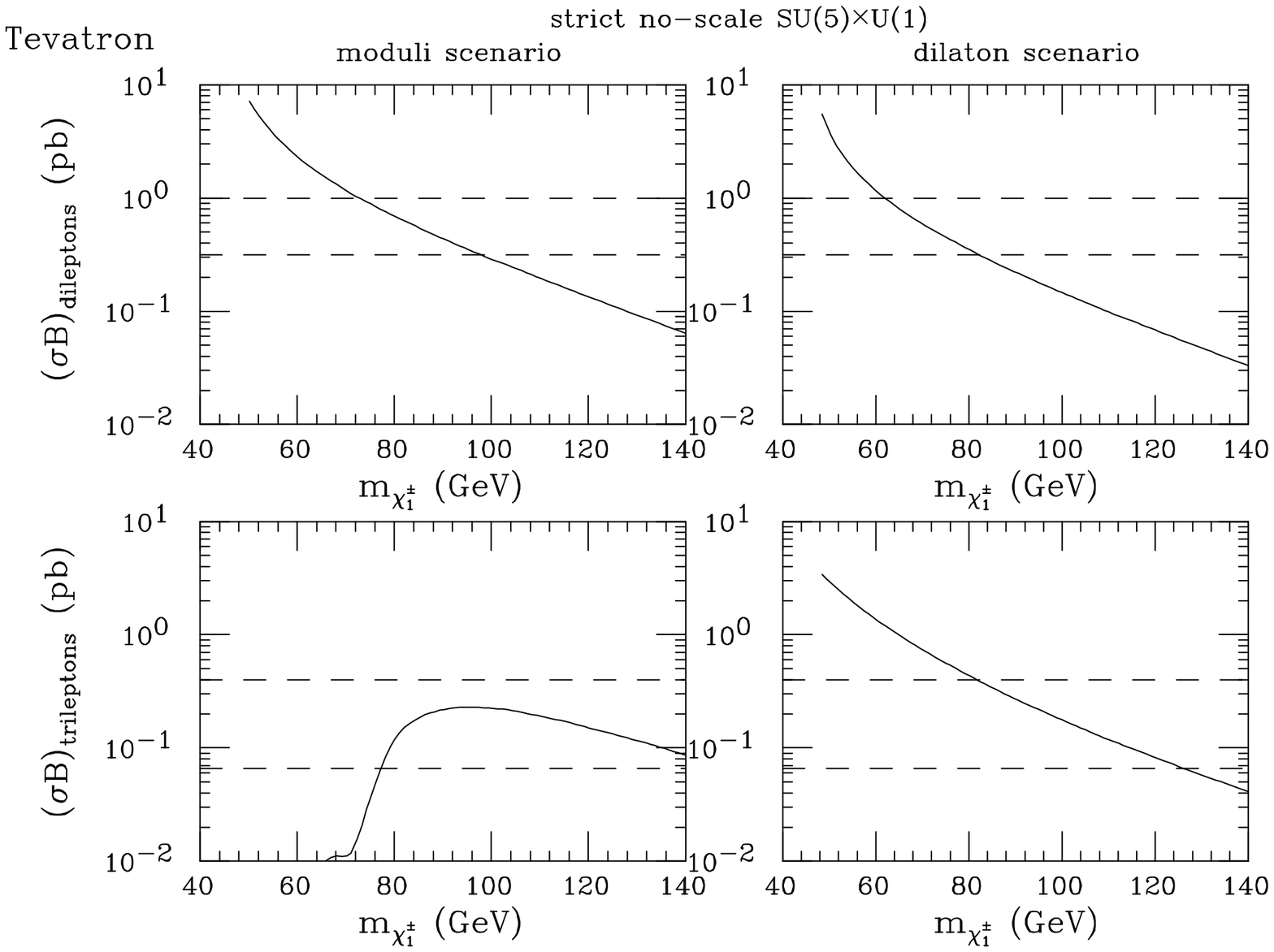}
\vspace{-2cm}
\caption{Dilepton and trilepton production at the Tevatron in one-parameter
$SU(5)\times U(1)$ supergravity. The upper (lower) dashed line represents the
estimated sensitivity for $100\,{\rm pb}^{-1}\,(1\,{\rm fb}^{-1})$. Note the
complementary nature of the two signals.}
\label{di-tri}
\end{figure}
%\clearpage

\subsection{Collider indirect tests}
$\bullet$ Determination of $B(Z\to\chi^0_1\chi^0_2)\lsim10^{-4}$ at LEP
constrains the $\chi^0_1,\chi^0_2$ masses. Using the relation
$m_{\chi^\pm_1}\approx m_{\chi^0_2}\approx 2m_{\chi^0_1}$, we find that
$m_{\chi^\pm_1}\gsim50\GeV$ is required, unless $\tan\beta\lsim2$, in which
case the $Z\chi^0_1\chi^0_2$ coupling is very small.\\
$\bullet$ Global fits to the electroweak data yield
$m_t=160^{+11}_{-12} {}^{+6}_{-5}\GeV$ in the context of supersymmetric models
with a light Higgs boson ($m_h=(60-150)\GeV$ gives the second error). This
program may constrain the supersymmetric sector too~\cite{EFL}.\\
$\bullet$ The averaged LEP measurement of $R_b=\Gamma(Z\to b\bar
b)/\Gamma(Z\to{\rm hadrons})$ is about $2\sigma$ higher than the Standard Model
prediction \cite{LEPC}. Moreover, supergravity-like contributions to $R^{\rm
susy}_b$ make the total prediction for $R_b$ only slightly closer to the LEP
result~\cite{KKW}. However, a glance at the measurements that are being
averaged shows that this apparent discrepancy may eventually disappear.\\
$\bullet$ Measurement of $B(t\to\tilde t_1\chi^0_1)$ at the Tevatron would not
be as good as direct $\tilde t_1$ detection \cite{One}.

\subsection{Rare processes}
\subsubsection{$b\to s\gamma$}
This process has been observed at CLEO: $B(b\to s\gamma)=(1-4)\times10^{-4}$ at
95\% CL. There are various dominant contributions written schematically
as
\[B(b\to s\gamma)_{\rm SUSY}\propto \left[-|A_{\rm
SM}|-|A_{H^\pm}|\pm|A_{\chi^\pm_1}|-|C|\right]^2\]
Because of the chargino contribution, $B(b\to s\gamma)_{\rm SUSY}$ can
be larger or smaller than the Standard Model prediction. QCD corrections
are large, and a complete two-loop calculation is required to settle the
scale ($Q$) uncertainty, which can be estimated by
varying the scale $Q:{1\over2}m_b\to 2m_b$. Taking the theoretical calculations
at face value, large values of $\tan\beta$ appear disfavored, at least for one
sign of $\mu$ \cite{LargeTanB}.
\subsubsection{$(g-2)_\mu$}
The last measurement of $a_\mu={1\over2}(g-2)_\mu$, \ie,
$a^{exp}_\mu=1165923\,(8.5)\times10^{-9}$ dates back to 1970. The latest
Standard Model prediction is $a^{\rm  SM}_\mu=1165919.20\,(1.76)\times10^{-9}$.
Thus, at the 95\%CL we can tolerate a new contribution in the range
$-13.2\times10^{-9}<a^{susy}_\mu<20.8\times10^{-9}$. In 1996 the new
Brookhaven E821 experiment will start running with an expected sensitivity of
$0.4\times10^{-9}$, which is adequate to observe the electroweak contribution
($\sim2\times10^{-9}$). A difficulty in this test is that the hadronic
uncertainty in the Standard Model prediction is of a similar magnitude,
although experiments at Novosibirsk should reduce this uncertainty
considerably. Such impasse should be of no consequence to tests of
supersymmetric models, since their predictions can easily exceed the present
allowed interval \cite{g-2}.

\subsection{Proton decay}
The prediction for the proton lifetime depends strongly on the GUT group.
In the case of $SU(5)$, dimension-six operators are mediated by $X,Y$ gauge
bosons and the largest mode is $p\to e^+\pi^0$, with
$\tau_p\sim3.3\times10^{35}(M_U/10^{16})^4$, which is basically
unobservable, although it implies $M_U\gsim10^{15}\GeV$. More important are
the dimension-five operators, which need to be dressed by a chargino loop. The
largest contribution comes from CKM mixing with the second generation, and the
largest mode is $p\to\bar\nu_{\mu,\tau}K^+$. Schematically one has
\[
\tau(p\to\bar\nu_{\mu,\tau}K^+)\sim\left|M_H\sin2\beta{1\over f}
{1\over 1+y^{tK}}\right|^2
\]
where $M_H$ is the Higgs triplet mass, $f$ is the one-loop dressing function
($f\sim m_{\chi^\pm_1}/m^2_{\tilde q}$, \ie, heavy squarks, light charginos are
preferred), and $1+y^{tK}$ gives the ratio of 3rd/2nd generation contributions
to the dressing. Requiring $m_{\tilde q,\tilde g}\lsim1\TeV$, one finds for
$M_H<3M_U\,(10M_U)$: $\tan\beta\lsim6\,(10)$ and $\xi_0\gsim6\,(4)$
\cite{AN+LNP}. Moreover, the relic density cosmological constraint requires
$m_{\chi^\pm_1}\lsim120\GeV$, since one needs to be near the $\chi\chi\to
h,Z\to f\bar f$ annihilation poles \cite{ANcosm}.

\subsection{Dark matter}
\subsubsection{Cosmology}
Conservation of $R$-parity in supergravity models implies the existence of
a stable lightest supersymmetric particle (LSP), which corresponds to the
lightest neutralino ($\chi$). The relic abundance $\Omega_\chi h^2$ can be
computed from the pair-annihilation amplitude, and depends on all supersymmetry
parameters, with more accurate methods needed near $s$-channel poles. If one
requires a Universe at least
10 Gyr old, one must demand $\Omega_\chi h^2<1$ (in a pure cold dark matter
universe). One can also consider various structure formation models, such as
the cold plus hot dark matter model (CHDM) with $\Omega_\chi\sim0.7$ and
$\Omega_\nu\sim0.3$, or the cosmological constant model (CC) with
$\Omega_\chi\sim0.2$ and $\Omega_\Lambda\sim0.8$. With a given value for the
Hubble parameter, such as $h=0.80\pm0.17$ from the Hubble Space Telescope, one
can obtain ranges for $\Omega_\chi h^2$ which can be contrasted with the
predictions of specific supergravity models (see Fig.~\ref{dm} \cite{One}).

\begin{figure}[t]
\vspace{5in}
\includegraphics{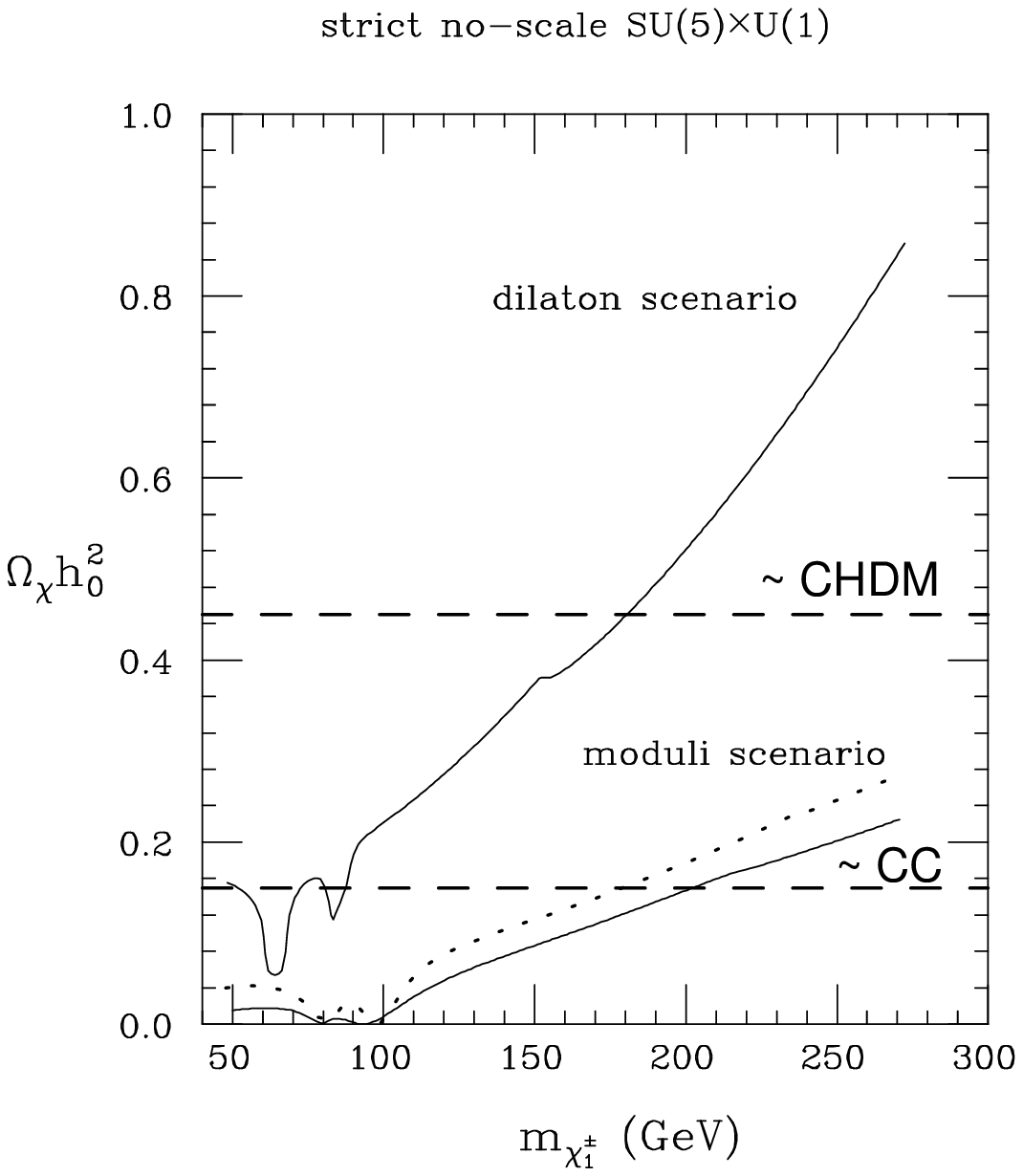}
\includegraphics{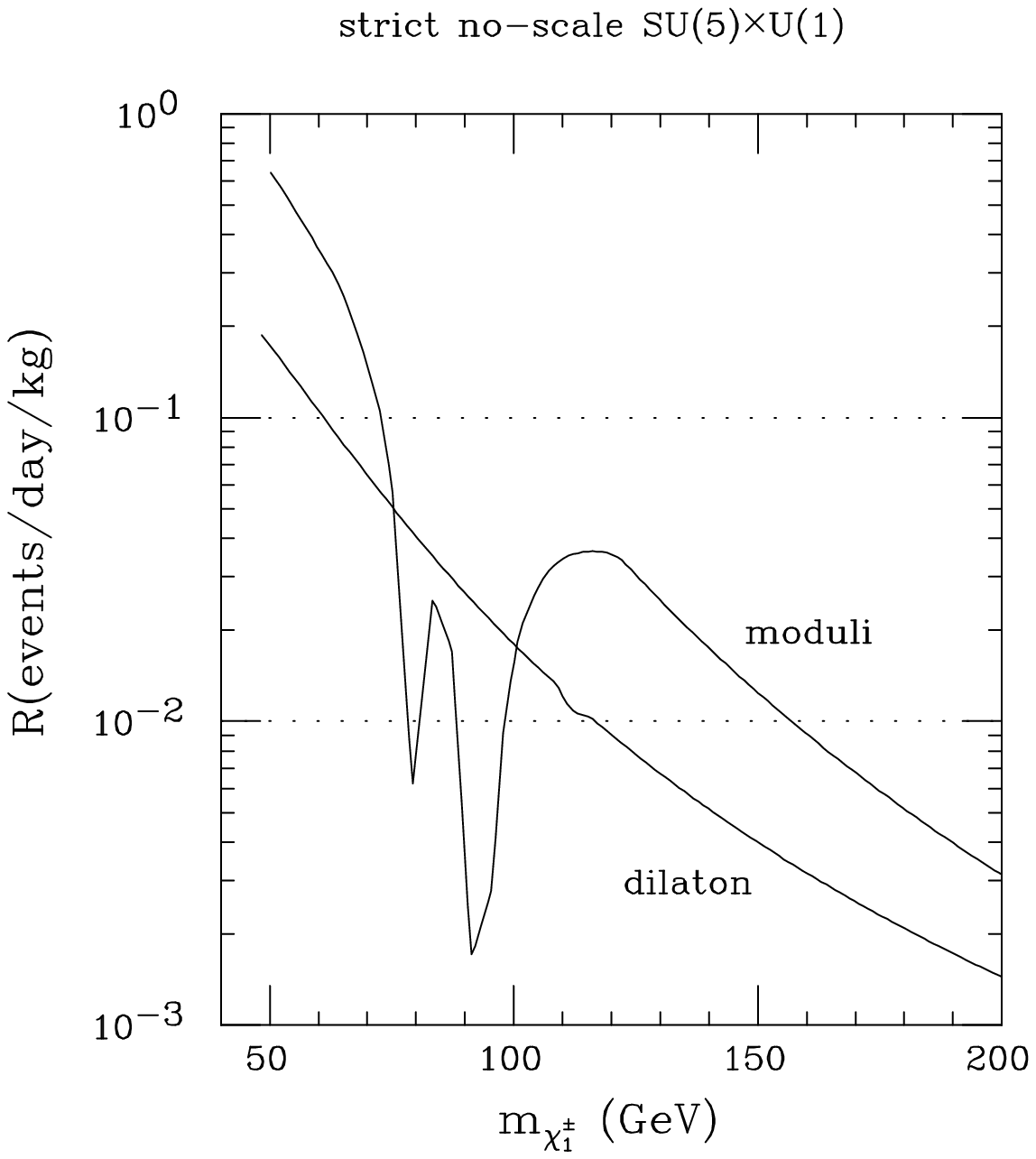}
\vspace{-5.25cm}
\caption{Relic abundance of neutralinos in one-parameter $SU(5)\times U(1)$
supergravity. Needed amounts in two cosmological models (CHDM and CC) can
discriminate between supergravity models. Also shown the direct detection rates
in the Ge detector. The dashed lines indicate expected near- and long-term
sensitivities.}
\label{dm}
\end{figure}

\subsubsection{Direct detection: cryogenic detectors}
Neutralino scattering off nuclei in cryogenic detectors is a promising avenue
to detect the LSP, especially in view of the new generation of detectors coming
on line (the Ge detector at Stanford) or in development stage \cite{Akerib}. It
is important to keep in mind that since $\Omega_{\rm halo}\gsim0.1$, one
requires $\Omega_\chi h^2\gsim0.05$ to get the ``full load" of LSPs, otherwise
the neutralino halo density should be scaled down. In Fig.~\ref{dm} we show
typical rates ($R$) for the Ge detector \cite{lspd}, where the depletion of
LSPs near the $Z$ and $h$ poles is evident. We also note that experimentally
excluded values for $B(b\to s\gamma)$ happen sometimes for interestingly large
values of $R$ \cite{BDN}.

\subsubsection{Indirect detection: Neutrino telescopes}
Neutralinos can also be captured by the Sun or Earth and then pair annihilate
producing eventually high-energy neutrinos which can be detected in a growing
number of under ground/water/ice detectors such as Kamiokande, MACRO, Dumand,
Amanda, Nestor. This detection technique is competitive with the direct one
\cite{Wells}.
\vspace{-0.1cm}
\section{Conclusions}
Given a few-parameter supergravity model one can combine all the above
constraints to determine the still-allowed region of parameter space to be
further explored at the Tevatron, LEPII, and the LHC, or through the rare
processes and indirect detection methods surveyed above \cite{Easpects,One}. We
conclude by  remarking that supergravity models are well motivated
few-parameter low-energy supersymmetric models, in which experimental
constraints can be enforced and testable predictions can be worked out in
detail. Experimentalists like them because they are ``easy targets" to kill or
severely damage. The procedure described above will need to be applied to the
better motivated models for the soft breaking parameters which are just
emerging from superstring model building. A lot of surprises may be in store.

\end{document}